\documentclass[conference]{IEEEtran}
\IEEEoverridecommandlockouts
\usepackage{cite}
\usepackage{amsmath,amssymb,amsfonts}
\usepackage{algorithmic}
\usepackage{graphicx}
\usepackage{textcomp}
\usepackage{xcolor}
\usepackage{algorithm}
\usepackage{booktabs}
\usepackage{svg}
\usepackage{siunitx}
\def\BibTeX{{\rm B\kern-.05em{\sc i\kern-.025em b}\kern-.08em
    T\kern-.1667em\lower.7ex\hbox{E}\kern-.125emX}}
\begin{document}

\title{Self-Supervised and Topological Signal-Quality Assessment for Any PPG Device\\
\thanks{This work was funded in part by NSF CHEST industrial sponsors.}
}

% \author{\IEEEauthorblockN{Wei Shao}
% \IEEEauthorblockA{\textit{Department of Computer Science} \\
% \textit{University of California, Davis}\\
% Davis, USA \\
% wayshao@ucdavis.edu}
% \and
% \IEEEauthorblockN{Ruoyu Zhang}
% \IEEEauthorblockA{\textit{Department of Electrical and Computer Engineering} \\
% \textit{University of California, Davis}\\
% Davis, USA \\
% ryuzhang@ucdavis.edu}
% \and
% \IEEEauthorblockN{Zequan Liang}
% \IEEEauthorblockA{\textit{Department of Computer Science} \\
% \textit{University of California, Davis}\\
% Davis, USA \\
% zqliang@ucdavis.edu}
% \and
% \IEEEauthorblockN{Ehsan Kourkchi}
% \IEEEauthorblockA{\textit{Department of Electrical and Computer Engineering} \\
% \textit{University of California, Davis}\\
% Davis, USA \\
% ekay@ucdavis.edu}
% \and
% \IEEEauthorblockN{Setareh Rafatirad}
% \IEEEauthorblockA{\textit{Department of Computer Science} \\
% \textit{University of California, Davis}\\
% Davis, USA \\
% srafatirad@ucdavis.edu}
% \and
% \IEEEauthorblockN{Houman Homayoun}
% \IEEEauthorblockA{\textit{Department of Electrical and Computer Engineerin} \\
% \textit{University of California, Davis}\\
% Davis, USA \\
% hhomayoun@ucdavis.edu}
% }

\author{
    Wei Shao$^{1}$, Ruoyu Zhang$^{2}$, Zequan Liang$^{1}$, Ehsan Kourkchi$^{2}$, Setareh Rafatirad$^{1}$, \\ %
    and Houman Homayoun$^{2}$ \\ \\%
    $^{1}$Department of Computer Science, University of California, Davis, Davis, CA, U.S.A.\\ %
    $^{2}$Department of Electrical and Computer Engineering, University of California, Davis, Davis, CA, U.S.A.\\%
    Email: \{wayshao, ryuzhang, zqliang, ekay, srafatirad, hhomayoun\}@ucdavis.edu\\%
}

\maketitle

%%%%%%%%%%%%%%%%%%%%%%%%%%%%%%%%%%%%%%%%%%%%%%%%%%%%%%%%%%%%%%%%%%%%%%%%
\begin{abstract}
Wearable photoplethysmography (PPG) is embedded in billions of devices, yet its optical waveform is easily corrupted by motion, perfusion loss, and ambient light—jeopardizing downstream cardiometric analytics. Existing signal-quality assessment (SQA) methods rely either on brittle heuristics or on data-hungry supervised models. We introduce the first fully unsupervised SQA pipeline for wrist PPG. Stage 1 trains a contrastive 1-D ResNet-18 on 276 h of raw, unlabeled data from heterogeneous sources (varying in device and sampling frequency), yielding optical-emitter– and motion-invariant embeddings (i.e., the learned representation is stable across differences in LED wavelength, drive intensity, and device optics, as well as wrist motion). Stage 2 converts each 512-D encoder embedding into a 4-D topological signature via persistent homology (PH) and clusters these signatures with HDBSCAN. To produce a binary signal-quality index (SQI), the acceptable PPG signals are represented by the densest cluster while the remaining clusters are assumed to mainly contain poor-quality PPG signals. Without re-tuning, the SQI attains Silhouette, Davies–Bouldin, and Calinski–Harabasz scores of 0.72, 0.34, and 6,173, respectively, on a stratified sample of 10,000 windows. In this study, we propose a hybrid self-supervised-learning–topological-data-analysis (SSL–TDA) framework that offers a drop-in, scalable, cross-device quality gate for PPG signals.
\end{abstract}

%%%%%%%%%%%%%%%%%%%%%%%%%%%%%%%%%%%%%%%%%%%%%%%%%%%%%%%%%%%%%%%%%%%%%%%%

%============================================================
\section{Introduction}\label{sec:intro}
%============================================================
Wearable photoplethysmography (PPG) underpins today’s cardiometric ecosystem—delivering heart rate, \mbox{SpO\textsubscript{2}}, respiration, and nascent cuff-less blood-pressure estimates in smartwatches, rings, and earbuds. Global shipments already exceed millions of units per year, generating petabyte-scale PPG streams. Yet the optical waveform is notoriously fragile: motion artifacts, ambient-light leakage, skin–sensor decoupling, and perfusion changes routinely degrade signal quality \cite{ragosta2017normal,price1993signals,wabnitz2015all}. Without timely filtering, downstream algorithms can yield grossly erroneous vitals, undermining user trust and clinical adoption.

Commercial firmware embeds hand-tuned signal-quality assessment (SQA) heuristics—thresholds on amplitude, template correlation, or derivative energy—engineered per LED wavelength and mechanical stack; a firmware update or strap relocation can break these rules. Supervised CNNs detect artifacts reliably \cite{pereira2019supervised}, but each hardware generation demands thousands of freshly labeled windows, rendering cross-device scaling impractical.

Wearables already store hundreds of hours of unlabeled wrist-PPG per user. Contrastive self-supervised learning (SSL) can harness this free data, but SSL alone does not output a human-interpretable signal-quality index (SQI). Conversely, topology-based descriptors capture waveform morphology in a few numbers, yet they have never been paired with modern deep encoders. Persistent homology (PH) has characterized cardiac periodicity and gait regularity \cite{graff2021persistent}; to our knowledge, we are the first to use PH as a morphology prior for wrist-PPG quality.

\begin{figure*}[t]
  \centering
  \includegraphics[width=\textwidth]{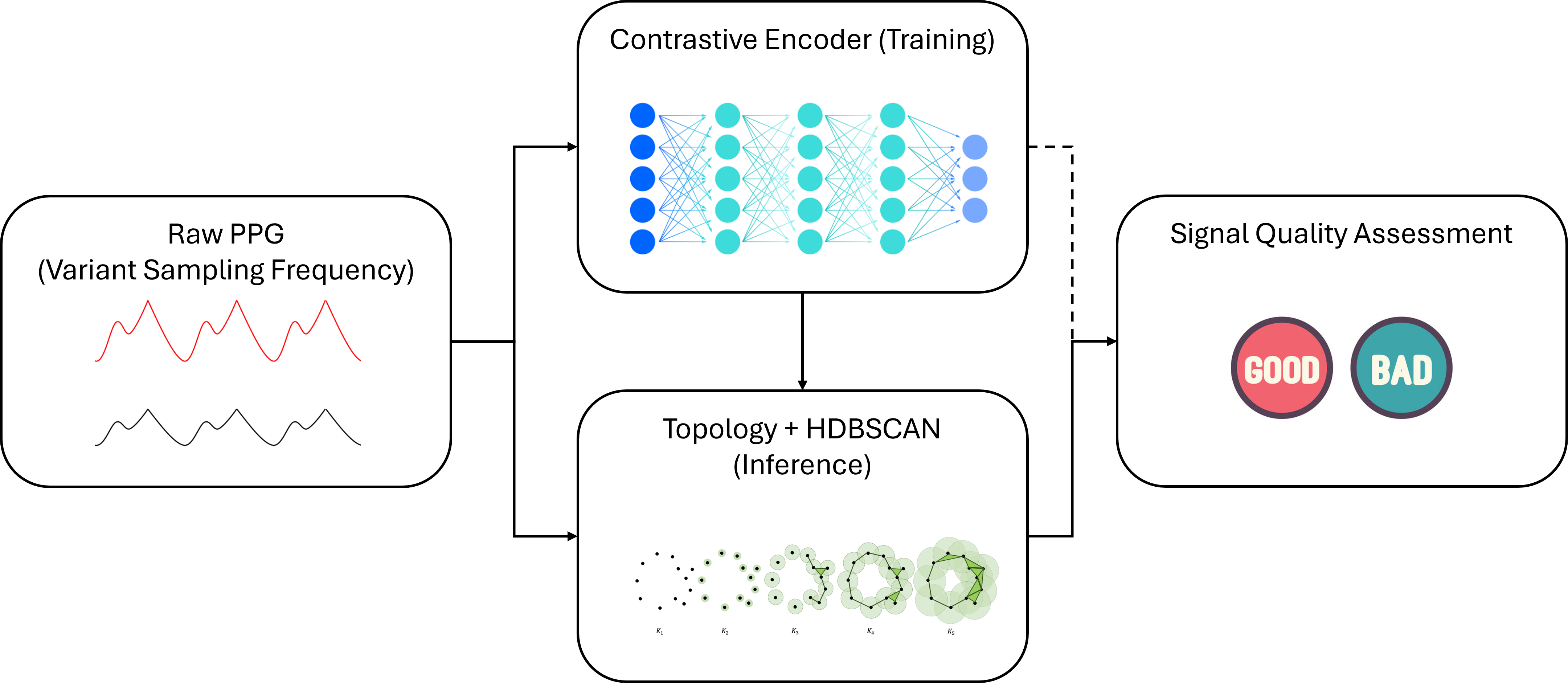}
  \caption{Proposed two-stage pipeline}
  \label{fig:pipeline}
\end{figure*}

We fuse SSL and topological data analysis (TDA) into the first fully unsupervised, device-agnostic SQA pipeline, shown in Fig. \ref{fig:pipeline}:

\begin{enumerate}
  \item \textbf{Contrastive representation learning:} 
  trains a contrastive encoder so that each 8\,s PPG window is mapped to a 512-dimensional embedding that is stable across device settings and motion artifacts.
  \item \textbf{Topology-driven quality discovery:}
  grounded in the invariance learned in Stage~1, we operate on these encoder-derived representations rather than on the raw waveforms. Specifically, we freeze the encoder, treat each 512-D embedding as a one-dimensional signal, compute a four-scalar PH signature of the embedding landscape and cluster these 4-D signatures with HDBSCAN; the largest dense cluster is labeled clean, while all other points are labeled poor.

  % Each 8\,s window is distilled into a four-scalar PH vector and clustered via HDBSCAN. The densest cluster is deemed \emph{clean}; everything else, \emph{poor}, yielding a binary SQI.
\end{enumerate}

% Table~\ref{tab:baseline_compare} contrasts existing SQA options on axes that matter for deployment. Heuristics (first row) are label-free and interpretable but brittle across hardware; supervised CNNs (second row) requires large per-device annotations and offer limited interpretability. Our self-supervised-learning–topological-data-analysis (SSL-TDA) approach remains label-free, portable across sampling rates/devices without re-tuning, and reduces each window to four interpretable scalars.

The key novelties are (i) the first SSL–TDA fusion for SQA, (ii) cross-device and -sampling rate portability without re-tuning, and (iii) an interpretable four-number signature enabling MCU-level inference.

% \begin{table}[tb]
% \centering
% \caption{Comparison of SQA approaches for PPG.}
% \begin{tabular}{lcccc}
% \toprule
% \textbf{Method} & \textbf{Label-free} & \textbf{Cross-device Portability} & \textbf{Interpretability} \\
% \midrule
% \cite{bashar2019atrial} & \textbf{Yes} & Low (re-tune thresholds) & \textbf{High} \\
% \cite{pereira2019supervised} & No & Low (retrain/relabel) & Low  \\
% SSL-TDA & \textbf{Yes} & \textbf{High} & \textbf{High}  \\
% \bottomrule
% \end{tabular}
% \label{tab:baseline_compare}
% \end{table}

%%%%%%%%%%%%%%%%%%%%%%%%%%%%%%%%%%%%%%%%%%%%%%%%%%%%%%%%%%%%%%%%%%%%%%%%
% \section{Related Work}\label{sec:related}

% \textbf{Heuristic SQI.}  Commercial wearables employ proprietary thresholds
% on peak height, template correlation, or accelerometer magnitude
% \cite{mclaughlin2020sqi}.  Heuristics are fast but brittle across devices.

% \textbf{Supervised SQI.}  CNNs and transformers trained on labelled windows
% achieve $>$90 \% accuracy \cite{banerjee2022ppgcnn}, yet require
% device-specific annotation and struggle to generalise across LED stacks.

%============================================================
\section{Background}\label{sec:background}
%============================================================

\subsection{Self-supervised learning for physiological signals}
Contrastive objectives such as SimCLR \cite{chen2020simple} and BYOL \cite{grill2020bootstrap} maximize agreement between two independently augmented views of the same instance; this technique outperforms autoencoders on ECG and PPG \cite{SSLECG,ding2024self} and on other biosignals by capturing invariance to amplitude scaling and temporal distortion with zero annotation effort.

\subsection{Persistent homology in time-series}
TDA quantifies the shape of data. Sublevel-set PH has characterized cardiac periodicity and gait regularity \cite{graff2021persistent,lamar2016persistent}. Clean, quasi-periodic PPG produces long-lived $H_{1}$ loops, whereas noisy windows do not, making PH an attractive unsupervised morphology cue. In addition, PH reduces encoder embeddings to morphology-aware scalars, adding an explicit morphological prior—capturing beat regularity versus artifact, and providing a compact and interpretable input to clustering, 

\subsection{Density-based clustering for quality discovery}
HDBSCAN extends DBSCAN with variable-density cluster extraction and explicit noise labeling \cite{mcinnes2017hdbscan}. It automatically chooses the number of clusters and handles non-Gaussian shapes—ideal for heterogeneous wrist data where artifacts are rare and scattered.

%======================================================================
\section{Methodology}\label{sec:method}
%======================================================================

% \subsection{Pipeline Overview}

% Our goal is to estimate PPG signal quality \emph{without} expert labels. Accordingly, we build a two-stage unsupervised pipeline:

% \begin{enumerate}
%   \item \textbf{Self-supervised representation learning.}  
%         % A self-supervised 1-D ResNet is trained with strong temporal
%         % augmentations so that each 8-s window containing the \emph{same} physiology
%         % map to nearby points in embedding space, regardless of amplitude,
%         % phase, or motion artifact.
%   \item \textbf{Topology-driven quality discovery.}  
%         % Persistent-homology features are extracted from each 8-s window and
%         % clustered with HDBSCAN.  
%         % The largest, densest cluster is interpreted as \textit{clean} PPG;
%         % all others are labelled \textit{poor} (noise), yielding a binary
%         % signal-quality index (SQI).
% \end{enumerate}

%----------------------------------------------------------------------
\subsection{Corpora and signal conditioning}
%----------------------------------------------------------------------
TABLE \ref{tab:data} lists the two datasets used in this study.

\begin{table}[b]
\caption{Unlabelled PPG corpora used for pipeline development.}
  \centering\small
  \begin{tabular}{lcccc}
    \toprule
    \textbf{Corpus} & \textbf{Site} & \textbf{Native $f_s$} & \textbf{Hours} & \textbf{LED}\\
    \midrule
    WildPPG~\cite{meier2024wildppg} & wrist & 128 Hz & 216 & green   \\
    We-Be~\cite{webe,webe-val}      & wrist &  25 Hz &  60 & green   \\
    \bottomrule
  \end{tabular}
  \label{tab:data}
\end{table}

\textbf{Why these datasets:}  
WildPPG offers long, mostly clean wrist recordings, whereas We-Be provides
lower-rate, motion-rich wrist data.  Joint training therefore encourages the
encoder to generalize across hardware and noise regimes.

\textbf{Signal conditioning:}  
A 0.5–8 Hz third-order, zero-phase Butterworth filter removes baseline
wander and LED noise. Traces are resampled to a common 25 Hz, $z$-scored,
and segmented into 8 s windows (200 samples, 50 \% overlap)

%----------------------------------------------------------------------
\subsection{Self-supervised representation learning (Contrastive Learning)}\label{ssec:ssl}
%----------------------------------------------------------------------
\paragraph{Loss function}  
The NT-Xent loss encourages invariance to amplitude and phase jitter—precisely the nuisance factors in wrist PPG—while requiring no annotations.

\paragraph{Encoder}  
A 1-D ResNet-18 processes 1\,$\times$\,200 inputs, followed by a projection
MLP (512$\rightarrow$512$\rightarrow$512). The output is $\ell_2$-normalized with
$\varepsilon=10^{-6}$.

\paragraph{Augmentation strategy}
Each view applies a deterministic band-pass filter, then draws two to four of the following transforms:
\begin{itemize}
  \item \textbf{Random crop (keep 50–70\%)}: packet loss, strap adjustment, transient motion gaps.
  \item \textbf{Time-warp ($\pm$3\%)}: natural heart-rate variability, slow sensor drift.
  \item \textbf{Jitter / Gaussian noise (1\% SD)}: sensor electronic noise, ambient light flicker.
  \item \textbf{Magnitude scaling ($\pm$5\%)}: LED drive-current fluctuations, skin perfusion changes.
  \item \textbf{Frequency dropout (narrowband removal)}: ambient light interference, missing harmonics.
  \item \textbf{Circular shift ($\pm$1\,s) and polarity inversion}: strap orientation errors, polarity mismatches.
  \item \textbf{Segment blackout (10–40 samples)}: short motion spikes (e.g., hand taps).
\end{itemize}
Empirical studies confirm that augmentations including jitter (Gaussian noise), scaling, time-warp, and polarity inversion reliably mimic motion, noise, and perfusion artifacts in contrastive learning for ECG/PPG signals~\cite{ding2024self,soltanieh2022analysis}.

\paragraph{Training details}  
By exploring hyperparameter tuning, we use the following: NT-Xent with $\tau=0.1$; batch size 512; AdamW (learning rate $2\times10^{-4}$, weight decay $10^{-4}$); 200 epochs; mixed precision.

% %----------------------------------------------------------------------
% \subsection{Topological signature}
% %----------------------------------------------------------------------

% Clean, quasi-periodic PPG traces generate pronounced one-dimensional loops
% ($H_1$) in their sublevel filtrations, whereas noisy windows do not.  For
% each 8 s window we compute persistent homology on a 1-D cubical complex
% (GUDHI) and retain four interpretable features:
% \[
% \bigl[n_{H_1},\ \Sigma H_1,\ \max H_0,\ \mathrm{mean}\,H_0\bigr]\in\mathbb{R}^4.
% \]

% %----------------------------------------------------------------------
% \subsection{Topological signature}
% %----------------------------------------------------------------------

% Clean, quasi-periodic PPG produces pronounced $H_1$ loops in sublevel filtrations, whereas noisy windows do not. We pass each 8\,s window through the trained encoder (frozen) and form a 1-D activation envelope from the penultimate conv feature map. PH on this envelope (GUDHI, 1-D cubical complex) yields four interpretable scalars—$n_{H_1}$, $\Sigma_{H_1}$, $\max_{H_0}$, and $\mathrm{mean}\,H_0$—which we feed to HDBSCAN in Stage~2.

%----------------------------------------------------------------------
\subsection{Topological signature}
%----------------------------------------------------------------------

We convert each 8\,s window into a 512-D embedding using the trained encoder (frozen). Interpreting this embedding as a one-dimensional scalar signal, we compute persistent homology on a 1-D cubical complex (GUDHI) and retain four interpretable features:
\[
\bigl[n_{H_1},\ \Sigma H_1,\ \max H_0,\ \mathrm{mean}\,H_0\bigr]\in\mathbb{R}^4.
\]

These four values summarize the structure in the embedding and are the inputs to HDBSCAN in Stage~2.

%----------------------------------------------------------------------
\subsection{Unsupervised quality discovery}
%----------------------------------------------------------------------
The 4-D persistence vectors are clustered with HDBSCAN. It adapts the number of clusters automatically, flags sparse points as noise, and handles non-Gaussian shapes—desirable for heterogeneous wrist PPG.

A binary SQI is assigned such that the largest non-noise cluster is deemed clean, and the remaining points (noise plus smaller clusters) are poor.

\subsection{Overall Performance}
Because the pipeline is label-free and device-agnostic by design, we evaluate structure quality using standard clustering validity scores (silhouette, Davies–Bouldin, Calinski–Harabasz) rather than supervised accuracy. These metrics capture separability and compactness of the discovered quality strata, which is appropriate when the objective is scalable, cross-device gating without annotation.

%======================================================================
\section{Evaluation}\label{sec:evaluation}
%======================================================================

\subsection{Encoder convergence}

The NT-Xent loss decreases smoothly from
$3.44\rightarrow0.95$ across 200\,epochs, while the mean cosine similarity
$\bigl(\overline{\cos}\bigr)$ between the two augmented views rises from
$0.67\rightarrow0.77$. The coupled evolution of
loss and cosine confirms that the encoder learns discriminative
directions rather than collapsing to a trivial representation.

%----------------------------------------------------------------------
\subsection{Quality of topological features}\label{ssec:ph_quality}
%----------------------------------------------------------------------

The $150\,000 \times 4$ persistence matrix exhibits a clear morphology
gradient: clean windows populate the high-$n_{H_1}$, high-$\sum H_1$ corner,
whereas noisy windows cluster near the origin.

\begin{figure}[tb]
  \centering
  \includegraphics[width=\columnwidth]{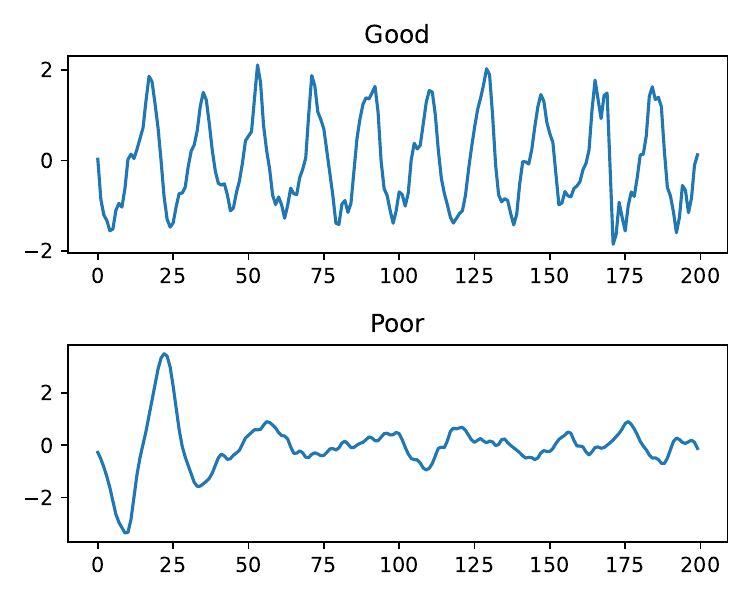}
  \caption{Visualization of the clustering result.}
  \label{fig:goodvspoor}
\end{figure}

Fig. \ref{fig:goodvspoor} shows a representative visualization of the
clustering outcome: the densest HDBSCAN cluster comprises 76\,\% of all
windows and corresponds to textbook pulsatile traces, explaining the great performance achieved by the SSL–TDA configuration.

%----------------------------------------------------------------------
\subsection{Ablation study}\label{ssec:ablation}
%----------------------------------------------------------------------
\begin{table}[tb]
\centering\small
\caption{Ablation on 10\,k windows}
\label{tab:ablation}
\begin{tabular}{lccc}
\toprule
\textbf{Configuration} & \textbf{Sil.($\uparrow$)} & \textbf{DB($\downarrow$)} & \textbf{CH($\uparrow$)} \\
\midrule
SSL+PH+HDBSCAN (SSL-TDA)          & \textbf{0.72} & \textbf{0.34} & 6\,173 \\
SSL+HDBSCAN (no PH)         & 0.05          & 4.60          & 29    \\
SSL+PH+$k$-means (no density)   & 0.39          & 0.89          & \textbf{7\,177} \\
SSL+$k$-means (SSL only)              & 0.01          & 7.33          & 141   \\
\bottomrule
\end{tabular}
\end{table}

TABLE \ref{tab:ablation} shows the ablation study results.
Removing the topological signature (SSL + HDBSCAN) collapses the
Silhouette from 0.72 to 0.05: in 512\,D, the contrastive embeddings form a
diffuse manifold that density-based clustering labels as almost homogeneous.
Conversely, retaining PH but swapping HDBSCAN for $k$-means halves the
Silhouette, showing that the density prior is also essential. The full
SSL–TDA fusion therefore yields the most compact and
well-separated clusters.

% \subsection{Unlabeled comparison with existing works}
% We assess convergent validity by comparing our SSL-TDA pipeline with two baselines on the same unlabeled corpus. For NeuroKit2\cite{makowski2021neurokit2}, we compute per-beat PPG template-matching scores and aggregate them to per-window values; for pyPPG\cite{goda2024pyppg}, we use its 0–1 template-matching SQI (the higher the better). To obtain binary outputs without labels, we prevalence-match thresholds so each method accepts the same fraction $q$ of windows (equal to our acceptance rate). Our SSL-TDA method agreed with NeuroKit2 on 82.22\% of windows and with pyPPG on 87.44\%.

\subsection{Unlabeled comparison with existing works}
We assess convergent validity by comparing our SSL-TDA pipeline with two baselines on the same unlabeled corpus. For NeuroKit2~\cite{makowski2021neurokit2}, we compute per-beat PPG template-matching scores and aggregate them to per-window values (median); for pyPPG~\cite{goda2024pyppg}, we use its 0-1 template-matching SQI. To obtain binary outputs without labels, we prevalence-match thresholds so each method accepts the same fraction $q=0.24$ of windows (equal to our acceptance rate). On $N=3600$ windows, it agreed with NeuroKit2 on 82.22\% and with pyPPG on 87.44\%.  These agreements indicate that our label-free method aligns with existing methods on most windows.

%============================================================
\section{Discussion}\label{sec:discussion}
%============================================================

% \paragraph{Why topology after contrastive learning?}
% Contrastive learning yields a \emph{feature space} in which pulses with the
% same underlying haemodynamics are close, yet it is agnostic to the absolute
% shape of the signal.  . It reduces each 512-D encoder embedding to four morphology-aware scalars, providing a compact and interpretable input to clustering. Table~II confirms that clustering on these PH features outperforms clustering directly on the full embeddings or on topology alone without SSL pre-training.
%   This hybrid design outperforms  
% (i) clustering directly in embedding space and  
% (ii) PH alone without SSL; both configurations achieve markedly lower
% silhouette scores.%

%  Table~II confirms that clustering on these PH features outperforms clustering directly on the full embeddings or on topology alone without SSL pre-training.

\paragraph{Dominant-cluster heuristic}
Our current implementation assumes that the largest and densest cluster discovered by HDBSCAN corresponds to physiologically clean PPG, while smaller or scattered clusters correspond to artifacts. This assumption holds in our corpora, where most windows contain usable signal, but it may break down in regimes dominated by noise. In such cases, the “clean = largest cluster” rule could invert. To mitigate this, one can (i) compute density ratios between the top two clusters and reject segments when the ratio falls below a threshold, (ii) weight clusters by intra-cluster persistence rather than point count, or (iii) use Bayesian non-parametric mixtures that relax the largest-cluster assumption. We note that the clustering framework can accommodate them without retraining the encoder.

% \paragraph{Why does the cosine similarity start high yet still work}
% In natural-image SimCLR the two random crops of a photo share only
% $\sim$10\,\% of pixels, so the initial view–view cosine is low
% ($\sim$0.1).  Here, both views receive the \emph{same} band-pass and contain
% the \emph{same} cardiac activity, differing only by mild time-warp,
% crop–pad, magnitude scaling, and related transforms.  Consequently, two
% randomly initialised encoder paths already observe highly correlated
% waveforms, and their embeddings begin with a non-trivial alignment:
% \(\overline{\cos}(z_1,z_2)\approx0.58\!-\!0.65\) across five random seeds.

% The critical check for representation \emph{collapse} is whether
% \(\overline{\cos}\) quickly approaches~1.0 \emph{without a simultaneous drop
% in loss}.  In our curves, the NT-Xent loss decreases monotonically
% ($3.44\rightarrow0.95$) while \(\overline{\cos}\) rises only modestly
% ($0.67\rightarrow0.77$), indicating that the encoder is learning new discriminative
% directions rather than mapping every input to an identical point.
% Empirically, windows from different subjects remain well separated in
% embedding space even at epoch 100, confirming that the higher baseline
% cosine is a benign consequence of domain-specific augmentations—not
% collapse.

\paragraph{Practical utility and downstream effects}
A natural question is whether the proposed SQA improves downstream analytics such as heart-rate estimation, rhythm classification, or biometric authentication. While we do not include full downstream validation here, prior studies have established that discarding poor-quality PPG segments reduces error rates in heart-rate monitoring and arrhythmia detection, and improves biometric authentication accuracy\cite{pereira2019supervised,chen2021clecg,shao2025know}. Our binary SQI removes roughly 24\% of windows in the We-Be dataset; in practice, this would filter the inputs to cardiometric pipelines so that algorithms operate on cleaner segments, reducing spurious beats and missed intervals. We position this work as a modular “quality gate” that can be inserted before such pipelines. A systematic evaluation of downstream benefits, such as pre/post SQI studies on shared labeled corpora—heart rate, rhythm classification, and biometrics, is an important direction for future work.

\paragraph{Beyond binary quality}
In this paper we report a binary SQI for clarity, assigning the largest dense cluster as clean and all others as poor. However, the clustering framework naturally produces multiple clusters and outlier scores, which could be mapped to finer-grained categories (e.g., clean / borderline / poor) or even a continuous quality index based on cluster density or silhouette distance. Such multi-level outputs may better match downstream applications (e.g., arrhythmia screening, where “borderline” segments should be flagged but not discarded).

\paragraph{Multi-modality}
Fusing accelerometer and PPG embeddings during contrastive pre-training may boost robustness to motion spikes that currently leak into the clean cluster. In addition, We-Be’s LED channels other than green were not exploited; multi-channel PH may further improve robustness. Finally, clinical validation against simultaneous ECG or invasive pressure would solidify the findings.

%============================================================
\section{Conclusion}\label{sec:conclusion}
%============================================================
We presented the first fully unsupervised two-stage pipeline that converts raw wrist-PPG into a binary SQI without device-specific thresholds or expert labels. Rather than optimizing supervised SQA accuracy, we prioritize scalability (no labels), portability (no device-specific re-tuning across 25–128\,Hz), and interpretability (four-scalar signature), positioning the method as a practical quality gate for diverse PPG devices, achieving Silhouette 0.72, Davies–Bouldin 0.34, and Calinski–Harabasz 6,173 on 276 h of heterogeneous data. Since it requires zero labels and no hardware calibration, the SSL–TDA framework can serve as a drop-in quality gate for any wrist-based PPG pipeline—paving the way for more reliable heart-rate, rhythm, and biometric-security analytics across the billions of wearables already in use.

% For future works, now we assume the clean cluster is densest; long recordings dominated by corruption may violate this.  Density-ratio tricks or Bayesian non-parametrics could relax the assumption. Fusing accelerometer and PPG embeddings during contrastive pre-training may boost robustness to motion spikes that currently leak into the clean cluster. We have not yet tested stability over weeks of wear or skin-tone diversity; a longitudinal, demographically balanced study is planned.
\bibliographystyle{IEEEtran}
\bibliography{ref.bib}
\vspace{12pt}
% \color{red}
% IEEE conference templates contain guidance text for composing and formatting conference papers. Please ensure that all template text is removed from your conference paper prior to submission to the conference. Failure to remove the template text from your paper may result in your paper not being published.

\end{document}